\documentclass[aps,prb,amssymb,showpacs,twocolumn]{revtex4-1}

\newcommand{\bk}{{\bf k}}
\newcommand{\bkp}{{{\bf k}^{\prime}}}
\newcommand{\bq}{{\bf q}}
\newcommand{\bh}{{\bf h}}
\newcommand{\bu}{{\bf u}}
\newcommand{\bp}{{\bf p}}
\newcommand{\uab}{u_{\alpha\beta}}
\newcommand{\uaa}{u_{\alpha\alpha}}

\newcommand{\br}{{\bf r}}
\newcommand{\bro}{{\bf r_0}}

\usepackage{graphics,graphicx,amsmath}

\usepackage{tabularx,float}
\usepackage{epsfig}

\usepackage{subfigure}

\usepackage{bm}
\usepackage{wasysym}

\usepackage{bm}
\usepackage{color}
\usepackage{verbatim}

\begin{document}

\bibliographystyle{apsrev4-1}

\date{\today}

\author{K. V. Zakharchenko, R. Rold\'an, A. Fasolino and M. I. Katsnelson}
\affiliation{Institute for Molecules and Materials, Radboud University
Nijmegen, Heyendaalseweg 135, 6525 AJ Nijmegen, The Netherlands}

\title{Self-Consistent Screening Approximation for Flexible Membranes: Application to Graphene}

\begin{abstract}
Crystalline membranes at finite temperatures have an anomalous behavior of the bending rigidity that makes them more rigid in the long wavelength limit. 
This issue is particularly relevant for applications of graphene in nano- and micro-electromechanical systems. 
We calculate numerically the height-height correlation function $G(q)$ of crystalline two-dimensional membranes, determining the renormalized bending rigidity,  in the range of wavevectors $q$ from $10^{-7}$ \AA$^{-1}$ till $10$ \AA$^{-1}$ in the self-consistent screening approximation (SCSA). For parameters appropriate to graphene,  the calculated correlation function agrees reasonably with the results of atomistic Monte Carlo simulations for this material within the range of $q$ from $10^{-2}$ \AA$^{-1}$ till $1$ \AA$^{-1}$. In the limit $q\rightarrow 0$ our data for the exponent $\eta$ of the renormalized bending rigidity $\kappa_R(q)\propto q^{-\eta}$ is compatible with the previously known analytical results for the SCSA $\eta\simeq 0.82$. However, this limit appears to be reached only for $q<10^{-5}$ \AA$^{-1}$ whereas at intermediate $q$  the behavior of $G(q)$ cannot be described by a single exponent. 

\end{abstract}


\maketitle

\section{Introduction}

A very active field in statistical mechanics and condensed matter physics is the study of interfaces and membranes. Physical membranes are two-dimensional surfaces embedded in three-dimensional space. In these systems, the interplay between the two-dimensional geometry and thermal fluctuations is at the origin of a number of unexpected behaviors, going from flat to glassy and tubular phases.\cite{NPW04} The stability of a flat 2D phase seems to be in contradiction with the Mermin-Wagner theorem,\cite{M68} which states the impossibility of long range order in two dimensions due to thermal fluctuations. This apparent contradiction became subject of great interest after the discovery of  graphene,  a single atom thick layer of carbon atoms,\cite{NF04,GN07,K07,G09,VKG10} which can be considered as the prototype of crystalline membranes. The stability of this material against crumpling, demonstrated even for free-standing samples,\cite{MR07,BG08} was proven to be related to the presence of intrinsic ripples.\cite{FLK07} Ripples and the mechanical properties of graphene have been subject of much recent theoretical work. \cite{KC08,GHD08,ZKF09,LF09,G09b,BH10,AART10}

The first attempt to study the anomalous elasticity in polymerized membranes was done by Nelson and Peliti using a simple one-loop self-consistent theory, without including any renormalization of the in-plane Lam\'e constants.\cite{NP87} They found an anomalous bending energy of the flat phase that for small wave vectors $q$ deviates from its constant value and acquires a power-law behavior for the effective bending rigidity $\kappa_R(q)\sim q^{-\eta}$ with $\eta=1$. The existence of anomalous elasticity was confirmed by an $\epsilon=4-D$ expansion, where $D$ is the membrane dimension.\cite{AL88} A step beyond was done by Le Doussal and Radzihovsky~\cite{DR92} who generalized to polymerized membranes the self-consistent screening approximation (SCSA) introduced by Bray~\cite{B74} to estimate the critical exponents of the O($n$) model in the large-$n$ limit. This approximation is exact when the co-dimension $d_c=d-D$ is going to infinity ($d$ being the dimension of the embedding space). In Ref.~\onlinecite{DR92} an approximate solution of the SCSA in the long wavelength limit was found, giving an exponent $\eta\approx 0.821$ for a $2D$ membrane in a $3D$ space.

Motivated by the relevance for graphene, several works have recently appeared studying the bending rigidity properties of 2D crystalline membranes. Mariani and von Oppen studied the one-loop correction to the bending rigidity due to the effective interaction between flexural phonons.\cite{MO08} More sophisticated methods as non-perturbative renormalization group (NPRG) have been used by Kownack and Mouhanna, who found an exponent of $\eta\approx0.85$,\cite{KM09} in good agreement with the SCSA results,\cite{DR92} and by Braghin and Hasselmann, who extended the analysis of Ref. \onlinecite{KM09} to finite momenta. Furthermore, the validity of SCSA has been recently checked by Gazit,\cite{G09c} who has applied the approximation to second order expansion in $1/d_c$  and found no significant deviation from the first order expansion. As a result, vertex corrections can be neglected during the calculation and SCSA seems to be applicable to crystalline membranes.

In this paper, we solve numerically the SCSA equations for the height-height correlation function $G(q)$ and calculate it in a wide  range of wavevectors $q$. In the long wavelengths limit $q \rightarrow 0$, our results for the exponent $\eta$  agree with the analytical solution of Le Doussal and Radzihovsky~\cite{DR92} but at larger $q$ the full solution has a more complex form that cannot be described by a single exponent.  Furthermore, we identify the lengthscale separating the harmonic behavior in the short wavelength limit, from the region where anharmonic coupling start to play an important role and the correlation function $G(q)$ is renormalized. We also compare the results of the numerical solution to Monte Carlo simulations of graphene based on the LCBOPII bond order potential.\cite{LF05} The two approximations reasonably agree, justifying the use of SCSA in the calculation of physical properties of graphene. 

\section{Method}

In this section we briefly review the SCSA for membranes.\cite{DR92,G09c} In the Monge representation, displacements of a $D$-dimensional membrane embedded in a $d$-dimensional space, are parametrized using a $D$-component phonon field $\bu$, and the out-of-plane height fluctuations by a $d_c=d-D$ dimensional field $\bh$. Therefore, if $\bro$ describes the position of a particle on the undistorted (flat) membrane, its configuration after the displacement due to perturbations will be given by the $d$-dimensional vector $\br=(\bro+\bu,\bh)$. Assuming an asymptotically flat geometry with small out-of-plane fluctuations, such that $\bu$ and $\bh$ are functions of $\bro$, the free energy takes the form:\cite{NPW04}
\begin{equation}\label{Eq:FreeEner}
F[\bu,\bh]=\frac{1}{2}\int d^D\br \left[\kappa (\nabla^2\bh)^2+2\mu \uab^2+\lambda \uaa^2 \right ],
\end{equation}
where the strain tensor $\uab$, to the lowest order in gradients of $\bu$ and $\bh$, reads
\begin{equation}\label{Eq:StrainTensor}
\uab\approx \frac{1}{2}(\partial_{\alpha}u_{\beta}+\partial_{\beta}u_{\alpha}+\partial_{\alpha}\bh\cdot\partial_{\beta}\bh),
\end{equation}
with $\alpha,\beta=1,...,D$. In Eq. (\ref{Eq:FreeEner}), $\kappa$, $\lambda$ and $\mu$ are the bending rigidity, the first Lam\'e constant, and the shear modulus respectively.\footnote{In our numerical calculations, we use the values valid for graphene at $T=300$~K, $\kappa\approx 1.1~\rm eV$, $\lambda\approx 2.4~\rm eV \AA^{-2}$ and $\mu\approx 9.95~\rm eV\AA^{-2}$. (See e. g. Ref.~\onlinecite{ZKF09}).} In the harmonic approximation, the last term of Eq.~(\ref{Eq:StrainTensor}) is neglected, leading to a decoupling of the bending ($h$) and stretching ($\bu$) modes. Eq.~(\ref{Eq:FreeEner}) provides a correct description of elastic free energy and height fluctuations of a membrane as long as the equilibrium phase is truly a flat phase. Once the phonons have been integrated out, the effective free energy can be expressed in terms of the Fourier components of the height fields
\begin{eqnarray}
F_{eff}[\bh]&=&\frac{1}{2}\int \frac{d^D\bq}{(2\pi)^D}\left [ \kappa q^4 |\bh_{\bq}|^2 +\frac{1}{4d_c}\int \frac{d^D\bk}{(2\pi)^D} \int \frac{d^D\bkp}{(2\pi)^D} \right .\nonumber\\
&\times&\left . R^{(D)}(\bk,\bkp,\bq)(\bh_{\bk}\cdot \bh_{\bq-\bk})(\bh_{\bkp}\cdot\bh_{-\bq-\bkp}) \right ],
\end{eqnarray}
where the effective four-point-coupling fourth-order tensor $R^{(D)}(\bk,\bkp,\bq)$ reads
\begin{eqnarray}
R^{(D)}(\bk,\bkp,\bq)&=&2\mu[\bk P^T(\bq)\bkp]^2\nonumber\\
&+&\frac{2\mu\lambda}{2\mu+\lambda}[\bk P^T(\bq)\bk][\bkp P^T(\bq)\bkp],
\end{eqnarray}
and $P^T_{\alpha\beta}(\bq)=(\delta_{\alpha\beta}-q_{\alpha}q_{\beta}/q^2)$ is the transverse projection operator. Notice that the interaction is completely separable for physical membranes ($D=2$ and $d=3$), allowing us to write:\cite{G09c} $R^{(2)}(\bk,\bkp,\bq)=2b_0[\hat{\bq}\times\bk]^2[\hat{\bq}\times\bkp]^2$, where $\hat{\bq}=\bq/q$ and $b_0=2\mu(\mu+\lambda)/(2\mu+\lambda)$.

Our aim is to calculate the correlation function 
\begin{equation}
\langle h_{\alpha}(-\bq)h_{\beta}(\bq)\rangle=\delta_{\alpha\beta}G(\bq),
\end{equation} 
with $G^{-1}(\bq)=\kappa q^4+\Sigma(\bq)$, where $\Sigma(\bq)$ is the self-energy and $G^{-1}_0(\bq)=\kappa q^4$ is the correlation function in the harmonic approximation. In the SCSA theory, the renormalized elasticity is determined through a $1/d_c$-expansion for the 2-point and 4-point correlation functions of $\bh$, that turns them into a closed self-consistent set of coupled integral equations for the self-energy $\Sigma(\bq)$. For physical membranes, the set of equations can be written as:~\cite{DR92}
\begin{eqnarray}
G^{-1}(\bq)&=&G_0^{-1}(\bq)+\Sigma(\bq)\label{Eq:SCEq-G}\\
\Sigma(\bq)&=&2\int \frac{d^2\bp}{(2\pi)^2}b(\bp)[\bq P^T(\bp)\bq]^2G(\bq-\bp)\label{Eq:SCEq-Sigma}\\
b(\bp)&=&\frac{b_0}{1+3b_0I(\bp)}\label{Eq:SCEq-b}\\
I(\bp)&=&\frac{1}{8}\int \frac{d^2\bq}{(2\pi)^2}q^2|\bp-\bq|^2G(\bq)G(\bp-\bq)\label{Eq:SCEq-I}
\end{eqnarray}
In Eq.~(\ref{Eq:SCEq-b}) the constants $\kappa$, $\lambda$ and $\mu$ appearing in $b_0$ are divided by $k_BT$, where $T$ is the temperature and $k_B$ the Boltzman constant. Eqs.~(\ref{Eq:SCEq-G})-(\ref{Eq:SCEq-I}) admit an analytic solution in the long wavelength limit, under the assumptions $G^{-1}(q)\approx \Sigma(q)\approx Z/q^{4-\eta}$, with $Z$ a non-universal amplitude, and $b(k)\approx 1/3I(k)$. The solution of such simplified system gives for the critical exponent $\eta=0.821$.~\cite{DR92} However, a full knowledge of the correlation function is lacking in this approach.

\section{Results and discussion}

\begin{figure}
\includegraphics[clip=true,width=0.9\linewidth]{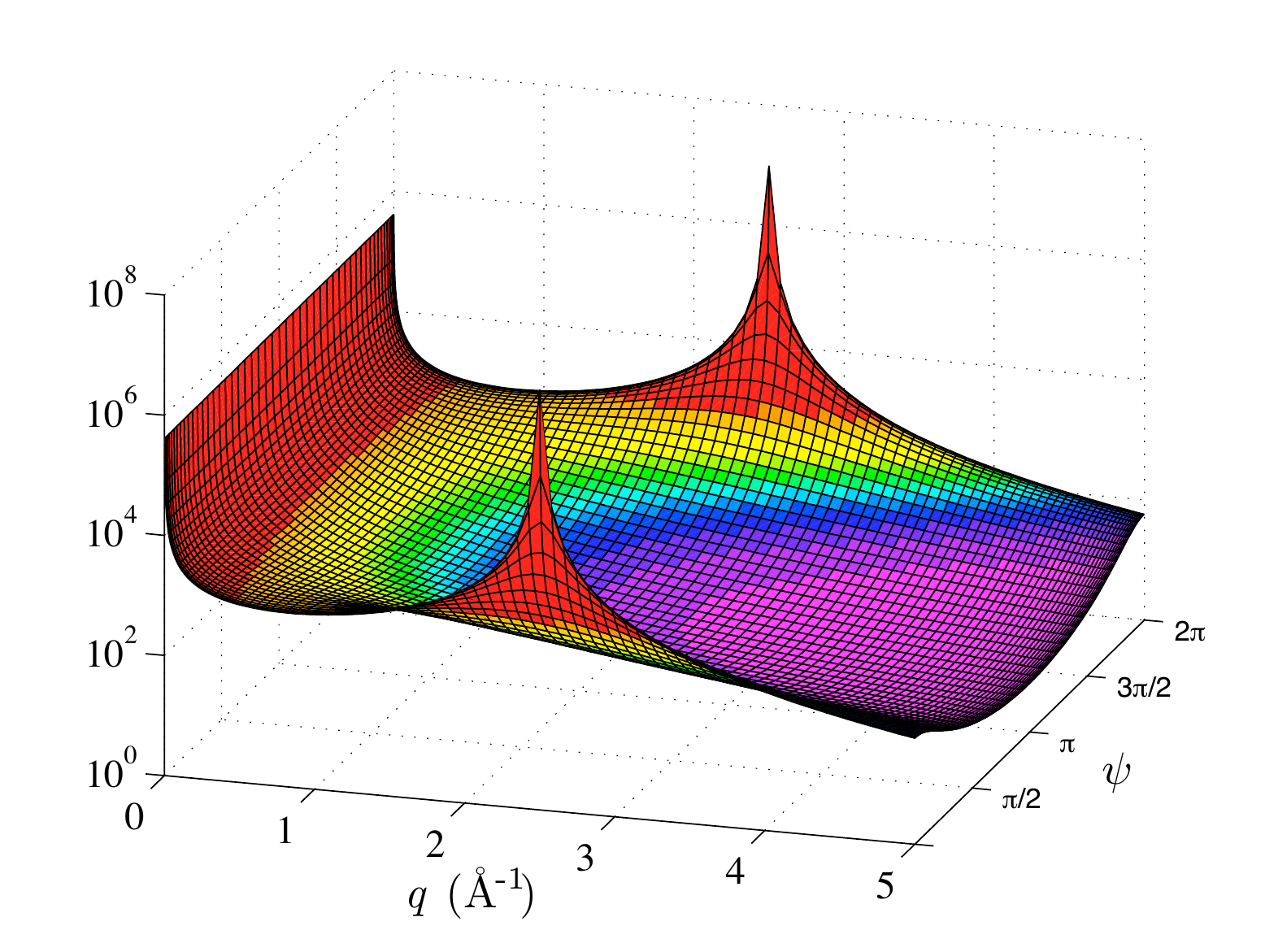}
\caption{(Color online) Integrand of Eq.~(\ref{Eq:SCEq-I}) for $p=2.5~ \rm \AA^{-1}$ and $\phi=0$. The UV cutoff in this figure has been taken, for illustrative reasons, to $q_{max}=5~\rm \AA^{-1}$.}
\label{Fig:I_q_subint_fig}
\end{figure}

In the following, we solve numerically the set of equations Eq.~(\ref{Eq:SCEq-G})--(\ref{Eq:SCEq-I}). The self-consistent cycle starts with the harmonic approximation $G(\bq)=G_0(\bq)$. From this, we compute Eq.~(\ref{Eq:SCEq-Sigma})--(\ref{Eq:SCEq-I}) and the obtained self-energy $\Sigma(\bq)$ is used to {\it dress} the new correlation function $G(\bq)$, which in turn allows us to start a new iteration. Taking into account that $G(\bq)$, $\Sigma(\bq)$, $b(\bq)$ and $I(\bq)$ depend only on the modules of the vector variables, it is natural to integrate Eqs.~(\ref{Eq:SCEq-G})-(\ref{Eq:SCEq-I}) in polar coordinates with the replacements $\bp \rightarrow (p, \phi)$ and $\bq \rightarrow (q, \psi)$. Moreover, further in this paper we will make no difference between $G(\bq)$ and $G(q)$. Thus, Eqs.~(\ref{Eq:SCEq-G})--(\ref{Eq:SCEq-I}) can be written as follows:

\begin{figure}[t]
  \centering
   \includegraphics[width=0.5\textwidth]{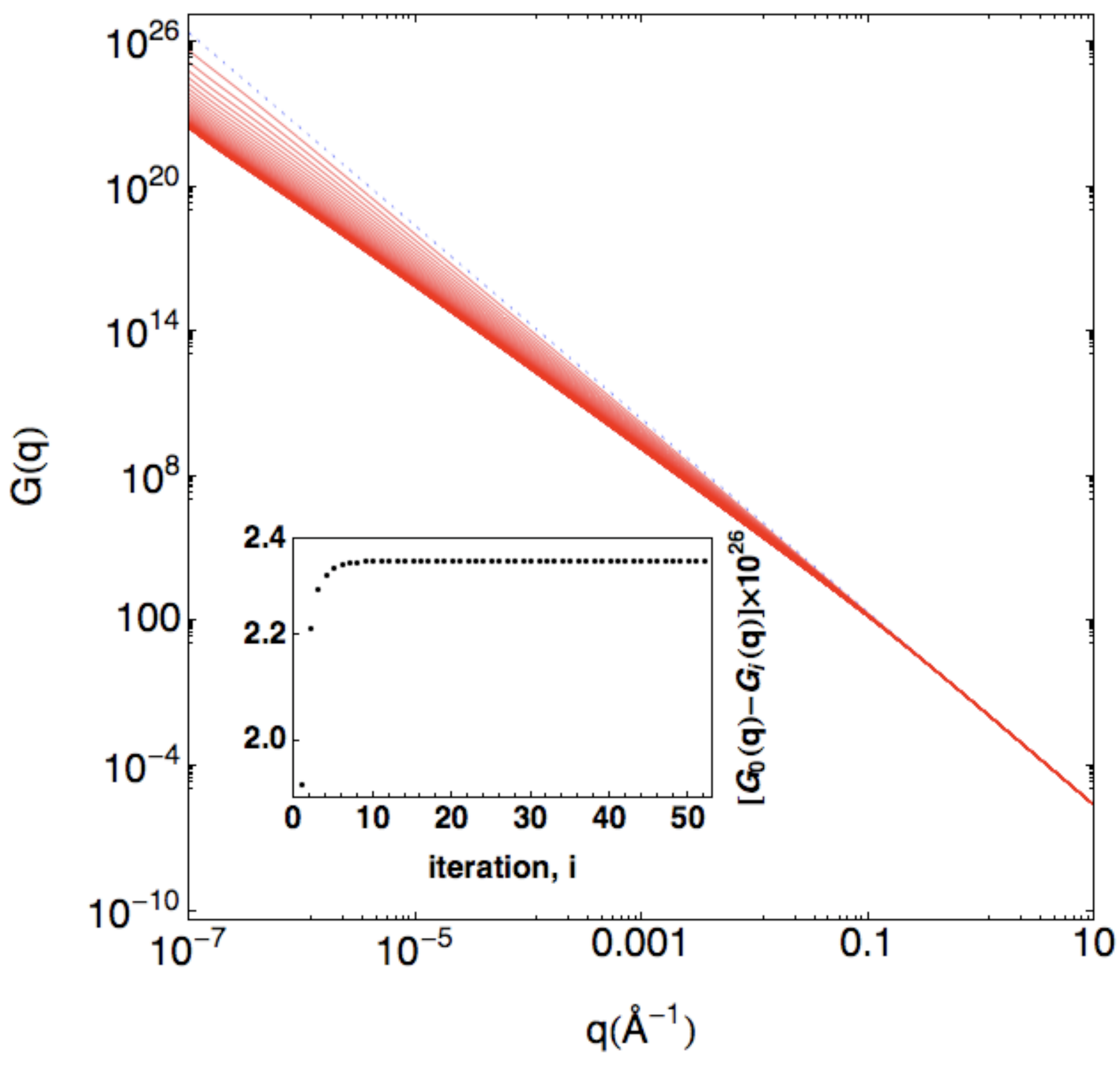}
 \caption{(Color online). Evolution of the calculated $G_i(q)$ for each iteration $i=1,...,51$ (red lines). $G_0(q)=\kappa/q^4$ is denoted by the dotted blue line. Inset: $G_0(q_0)-G_i(q_0)$ as a function of the iteration $i$ for $q_0=10^{-7}~\rm \AA^{-1}$, which shows how the solution converges after a few iterations.  }
  \label{Fig:Gconverg}
\end{figure}

\begin{gather}
G^{-1}(q, \psi) = G_0^{-1}(q, \psi) + \Sigma(q, \psi)\label{Eq:G_corr_2}\\
\Sigma(q, \psi) = \frac{1}{2 \pi^2} \int_0^{2\pi}\! d\phi\, \int_0^{q_{\rm max}}\!dp\, b(p, \phi) p q^4 \sin^4\!(\psi-\phi) \notag\\ \hfill \times G(\sqrt{q^2+p^2-2qp\cos(\psi-\phi)}, \psi-\phi)\label{Eq:sigma_of_q_2}\\
b(p, \phi) = \frac{b_0}{1+ 3 b_0 I(p, \phi)}\label{Eq:b_of_q_2}\\[1ex]
I(p, \phi) = \frac{1}{32 \pi^2} \int_0^{2\pi}\! d\psi \int_0^{q_{\rm max}}\! dq\, q^3 (q^2+p^2-2pq\cos(\phi-\psi))\notag\\ \hfill \times G(q, \psi) G( \sqrt{q^2+p^2-2pq\cos(\phi-\psi)}, \phi-\psi )\label{Eq:I_of_q_2}
\end{gather}
 
In the numerical implementation we have used a (hard) ultraviolet (UV) cutoff $q_{\rm max}$ in the radial integrals. Due to finite size effects, it is natural to consider an UV cutoff (which is of the order of the inverse lattice constant in crystalline membranes) and we have checked that the results are independent on this cutoff. We have checked that, in the relevant range,  the same results are obtained by multiplying $G(q)$ by a cutoff function $A(q)\sim e^{-\kappa (q/\Lambda)^4}$, where $\Lambda\simeq q_{\rm max}/5$ and we have used $q_{\rm max}=100~\rm \AA^{-1}$.

\begin{figure}[t]
  \centering
   \includegraphics[width=0.5\textwidth]{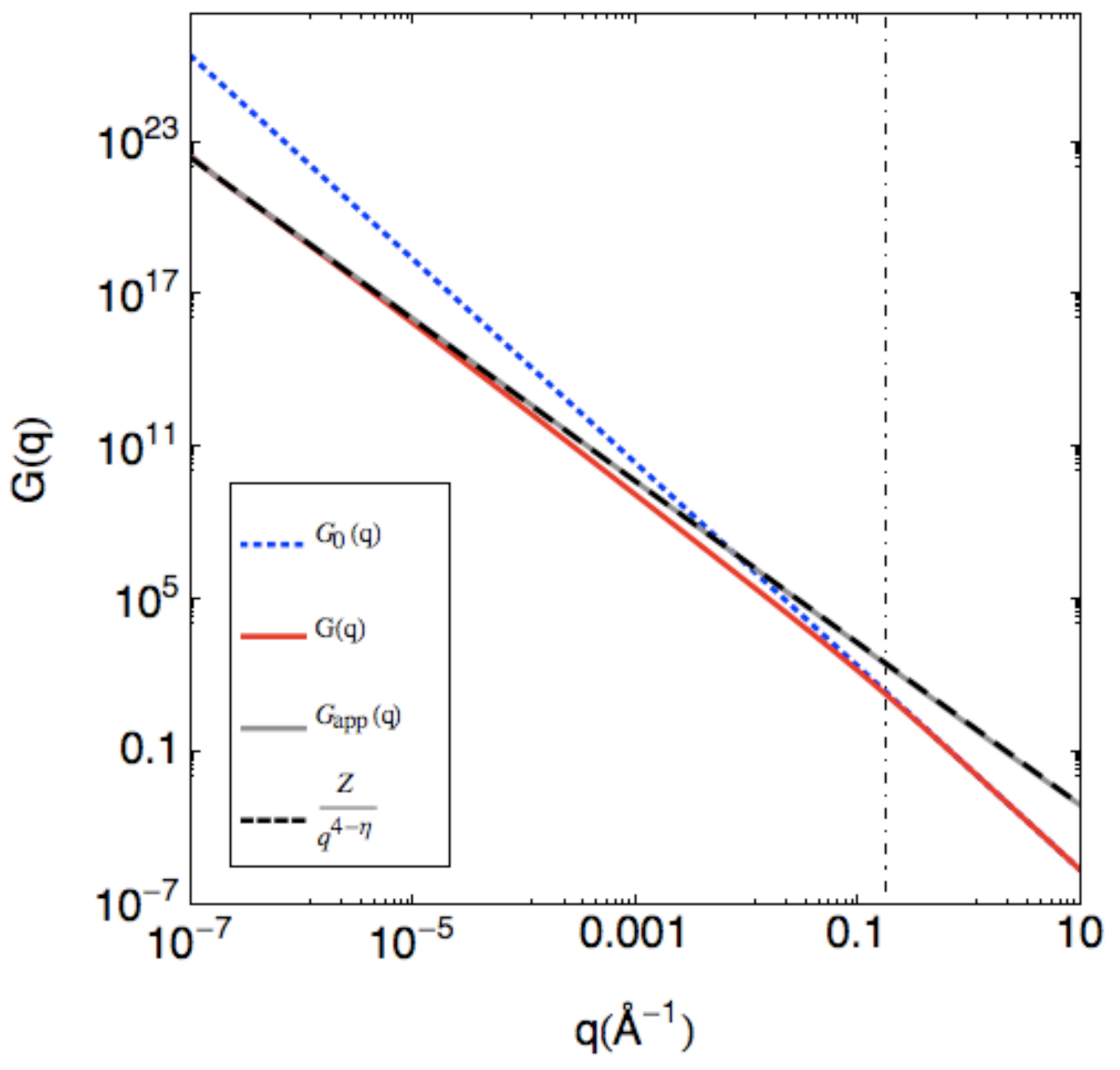}
 \caption{(Color online). Comparison of the unrenormalized correlation function in the harmonic approximation $G_0(q)$ (dotted blue line) to the solution of the SCSA equations $G(q)$ (red line) and the long-wavelength limit solution $G_{app}(q)$, using the approximations of Ref. \onlinecite{DR92} (gray line). The black dashed line is a fitting to the approximate solution $G(q)\approx Z/q^{4-\eta}$ choosing the parameters $\eta=0.821$ and $Z=1.2$. The vertical dot-dashed line indicates the wavevector $q^*\approx 0.18~\rm \AA^{-1}$ obtained from the Ginzburg criterion Eq. (\ref{Eq:q_crit}).}
  \label{Fig:Gcomp}
\end{figure}

The next difficulty is the divergence of the correlation function $G(q)$ in the infrared (IR) limit when $q\rightarrow 0$. As an example, Fig.~\ref{Fig:I_q_subint_fig} shows the integrand of Eq.~\ref{Eq:SCEq-I}, where the two divergences, for $q=0$ and $q=p$, can be seen. To avoid such IR divergence, we replace the function $G_0(q)=\kappa/q^4$ by $G_0(q)=\kappa/(q+\epsilon)^4$, where $\epsilon$ is a small parameter ($\epsilon = 10^{-46}~\rm \AA^{-1}$ in our numerical calculations).

Because of the power law behavior of the correlation function, it is extremely convenient to use a logarithmic grid for numerical evaluations. Therefore we discretize the momentum axis into points $q_i=ae^{d(i-1)}$, where $i$ is the index of the point in the grid of $q$, $a$ is the minimum value considered for the representation ($a=10^{-7}\rm \AA^{-1}$ in our calculations) and $\ell=\log(q_{\rm max}/a)/(N-1)$, where $q_{\rm max}$ is the UV cutoff and $N$ is the number of points in the grid of $q$.\footnote{For the numerical integration, we use the {\it nag\_quad\_md\_rect} algorithm of the NAG libraries, based on the HALF procedure.\cite{DR76,GM80}} In Fig.~\ref{Fig:Gconverg} we show the renormalized correlation function $G(q)$ after each of the first $51$ iterations. In general, convergence is very fast and achieved after about 20 iterations.

Our results are summarized in Fig.~\ref{Fig:Gcomp}. There we compare the bare (unrenormalized) correlation function $G_0(q)=1/\kappa q^{4}$ (dotted blue line) to the solution of the SCSA, $G(q)$ (red line). The important result is the value of the wavevector, $q_{c}\approx 0.1~\rm \AA^{-1}$, where $G(q)$ changes behavior from harmonic, where $G(q)\propto 1/q^4$ (for $q>q_{c}$), to non-harmonic, with $G(q)\propto 1/q^{4-\eta}$, for  $q<q_{c}$. The Ginzburg criterion~\cite{NPW04} gives an approximate value of the wavevector  $q^*$, and thus the spatial scale, $L^* = 2\pi/q^*$, at which anharmonic effects become dominant
\begin{equation}\label{Eq:q_crit}
q^*=\sqrt{\frac{3TK}{8\pi\kappa^2}},
\end{equation}
where $K$ is the $2D$ bulk modulus. For graphene, $K=12.4~\rm eV\cdot\AA^{-2}$ and $\kappa=1.1~\rm eV$ at room temperature ($T=300~\rm K$),~\cite{ZKF09} leading to $q^*\approx0.18~\rm \AA^{-1}$. This wavevector is represented by the vertical dotted-dashed line in Fig.~\ref{Fig:Gcomp}, and it is in good agreement with the SCSA results.

Furthermore, we have numerically solved the SCSA set of equations Eqs.~(\ref{Eq:SCEq-G})--(\ref{Eq:SCEq-I}) in the long wavelength approximation used by Le Doussal and Radzihovsky.~\cite{DR92} By taking $G^{-1}(q)\approx \Sigma(q)$ and $b(p)\approx 1/3I(p)$, we obtain the approximate solution shown by the green line in Fig.~\ref{Fig:Gcomp}, which is only valid in the long wavelength limit. Notice that both, the exact and the approximate solutions coincide for small wavevectors (i.e. in the limit $q \rightarrow 0$).

Finally, we have fitted this approximate solution to $G(q)\approx Z/q^{4-\eta}$, with $\eta=0.821$ and $Z=1.2$, as shown by the dashed black line ($q$ is expressed in $\rm \AA^{-1}$). The three results (exact numerical solution of the SCSA, approximate numerical solution and analytic approximation) coincides in the long wavelength limit, and corroborate the value given in Ref.~\onlinecite{DR92} for the critical exponent, $\eta=0.821$. We mention here that the above solution is robust as far as we start the first iteration from the harmonic approximation [$G_0(q)\sim q^{-4}$] or from any correlation function that diverges faster than $q^{-4+\eta_0}$ with $\eta_0\approx0.85$.

 \begin{figure}[t]
  \centering
   \subfigure[]{\label{Fig:GMCtotal}\includegraphics[width=0.53\textwidth]{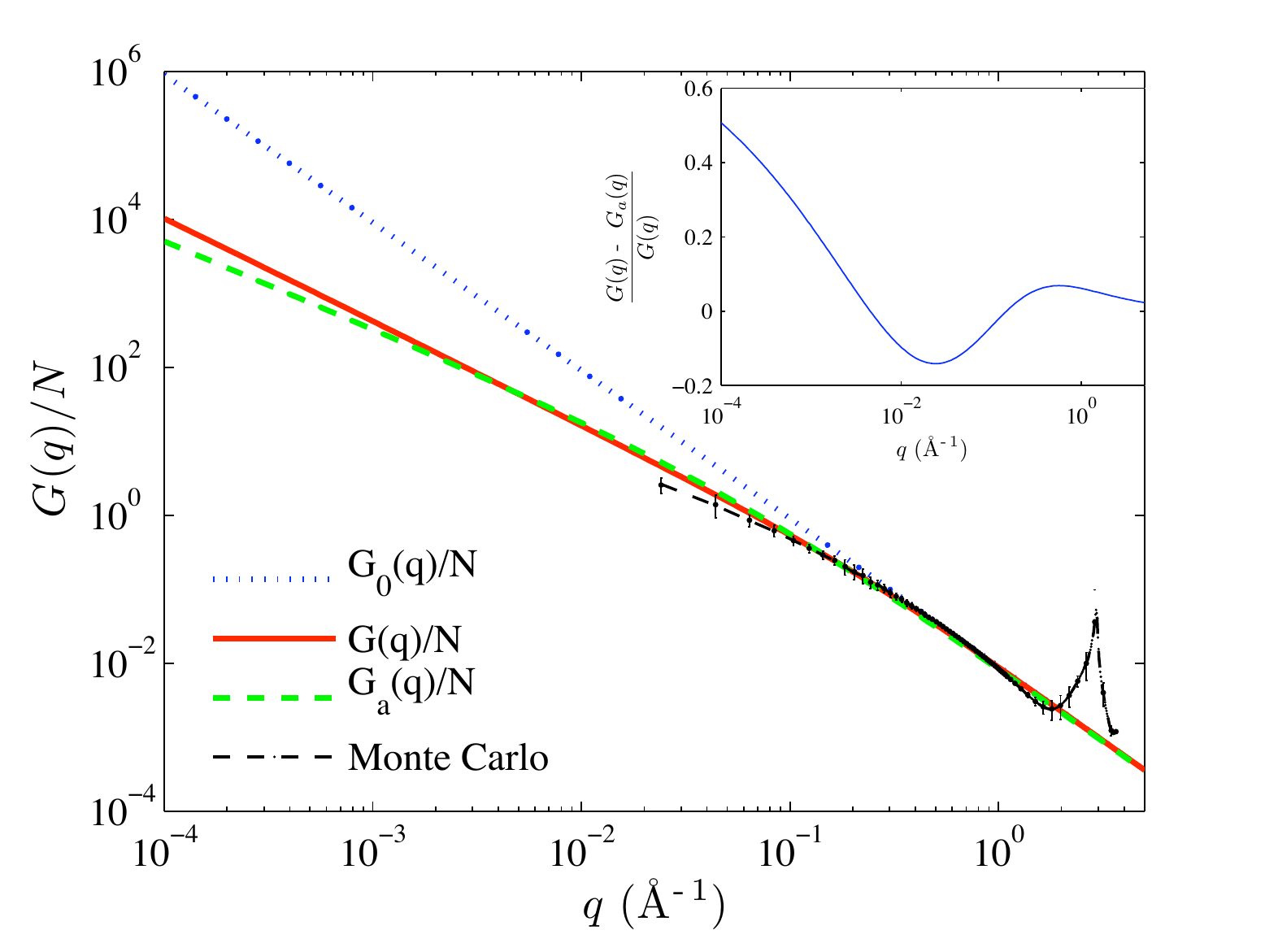}}
  \subfigure[]{\label{Fig:GMCzoom}\includegraphics[width=0.53\textwidth]{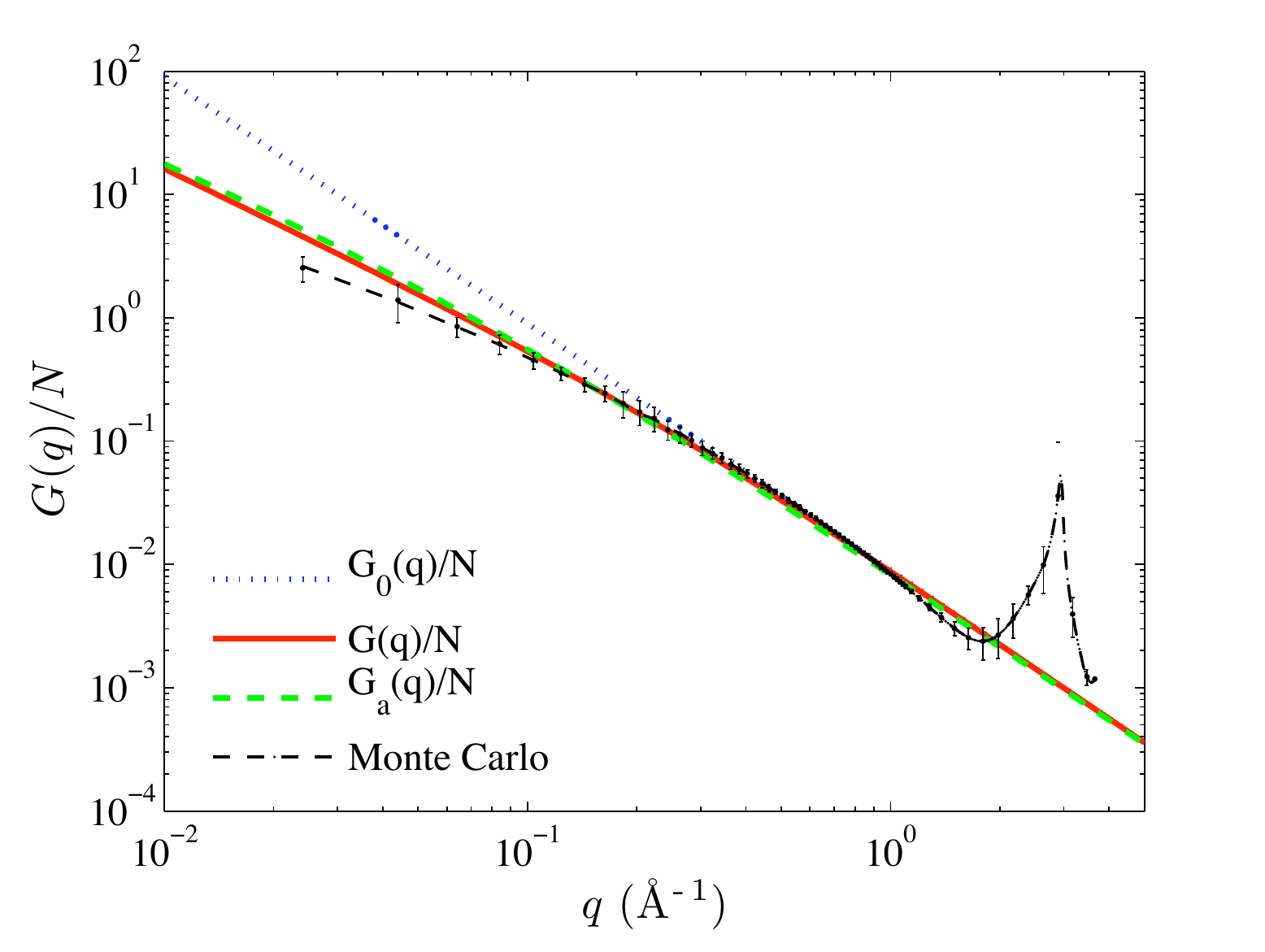}}
 \caption{(Color online). (a) Comparison of the unrenormalized correlation function in the harmonic approximation $G_0(q)$ (dotted blue line) to the solution of the SCSA equations $G(q)$ (red line) and the Monte Carlo data (black dot-dashed line). The dashed green line correspond to the approximation given by Eq. (\ref{Eq:Ga}). In the inset we show the deviation of the approximation $G_a(q)$ from the SCSA solution $G(q)$. (b) Zoom of Fig. \ref{Fig:GMCtotal} focusing on the comparison of $G(q)$ from SCSA to the Monte Carlo data.}
  \label{Fig:GMC}
\end{figure}

We also compare the solution of the SCSA system of equations with the correlation function $G(q)$ of graphene extracted from the Monte Carlo simulations presented in Ref.~\onlinecite{LF09}. For more details about the Monte Carlo calculation of the correlation function $G(q)$,  see Ref.~\onlinecite{ZLKF10}.
In Ref.~\onlinecite{LF09} the results for the correlation function found for two different model potentials were described by a power law with exponent $\eta=0.85$. 
 The Monte Carlo results are shown in Fig.~\ref{Fig:GMC} together with the solution of the SCSA system of equations and the unrenormalized correlation function $G_0(q)=1/\kappa q^4$.  In Fig.~\ref{Fig:GMC} we can see that $G(q)$ obtained from the SCSA equations agrees rather well with the Monte Carlo data in the range of $q$ accessible in atomistic calculations. An even better agreement with Mote Carlo data was found in Ref. \onlinecite{BH10}, where the height-height correlation function was computed using a more accurate approximation as the NPRG. However, notice that we do not use here any additional adjustable parameter when comparing to Monte Carlo data. Therefore, this justify the use of SCSA in the intermediate range of momenta.

Furthermore, we compare the results to the approximate correlation function $G_a(q)$, obtained from the effective Dyson equation\cite{FLK07}
\begin{equation}\label{Eq:Ga}
G_a^{-1}(q)=G_0^{-1}(q)+\Sigma(q),
\end{equation}
where $G_0(q)$ is the correlation function in the harmonic approximation
\begin{equation}
G_0(q)=\frac{N}{\kappa S_0 q^4},
\end{equation}
$N$ being the number of atoms of the sample and $S_0=L_xL_y/N$ the area per atom, and the self-energy is approximated by
\begin{equation}
\Sigma(q)=\frac{AS_0}{N}q^4\left ( \frac{q_0}{q} \right )^{\eta}
\end{equation}
where $q_0=2\pi\sqrt{K/\kappa}$ and $A$ an unknown numerical factor. The fitting of Eq. (\ref{Eq:Ga}) to the solution of the SCSA equations in the region $10^{-4}-1~\rm \AA^{-1}$ gives $A=0.3261$, as shown in Fig. \ref{Fig:GMC} by the dashed green line. In this fitting the exponent $\eta$ has been fixed to its long wavelength value, $\eta=0.82$. This approximation is a good interpolation function between the long- and short-wavelength regions, and it can be used to simplify the calculation of physical quantities that involve the renormalized correlation function. This range of wavevectors ($10^{-4}-1~\rm \AA^{-1}$) is relevant for discussing the scattering of electrons by ripples.\cite{KG08}

 \section{Conclusions}

In summary, we have studied numerically the self-consistent theory of polymerized membranes proposed in Ref. \onlinecite{DR92}. The critical exponent that we obtain in the long wavelength limit, $\eta\approx0.82$, coincides with the analytic approximation. In addition, we have calculated the correlation function $G(q)$ in the whole range of momenta and found good agreement with results of Monte Carlo calculations. We have also found the characteristic wavevector, $q_{c}\approx 0.1~\rm \AA^{-1}$, that separates the region of validity of the harmonic approximation (for $q\gtrsim q_{c}$) where $G(q)\propto q ^{-4}$, from the region where fluctuations lead to a considerable renormalization of the correlation function, and where $G(q)\propto q ^{-4+\eta}$. This value of $q_{c}$ is close to the one given by the Ginzburg criterion. From this wavevector, the exponent $\eta$ changes from zero (for $q\gtrsim q_{c}$) to 0.82 in the long wavelength limit. This limit is important when dealing with MEMS applications of graphene.\cite{BM07,LWKH08,BG08} The renormalization of the bending rigidity $\kappa\rightarrow \kappa_{R}(q)\sim q^{-\eta}$ should be taken into account, e. g., when calculating the eigen-frequencies of graphene membranes that would become $\omega(q)\propto \sqrt{\kappa_R(q)q^4} \propto q^{2-\eta/2} \simeq q^{1.6}$.

Our results show the importance of considering the renormalization of the bending rigidity. The good agreement between SCSA and Monte Carlo simulations for graphene can be seen as a proof that SCSA is a good approximation to account for the effect of corrugation in the physical properties of graphene.

 \begin{acknowledgments}
We thank Jan Los for discussions. This work is part of the research program of the 'Stichting voor Fundamenteel Onderzoek der Materie (FOM)', which is financially supported by the 'Nederlandse Organisatie voor Wetenschappelijk Onderzoek (NWO)'. We thank the EU-India FP-7 collaboration under MONAMI, and the Netherlands National Computing Facilities foundation (NCF).
 \end{acknowledgments}

\bibliography{BibliogrGrafeno}

\end{document}